\documentclass[a4paper,groupedaddress,superscriptaddress,twocolumn]{revtex4-1} 

\usepackage{graphicx}
\usepackage{physics}
\usepackage[usenames,dvipsnames]{xcolor}
\usepackage[normalem]{ulem}
\usepackage{rotating}
\usepackage{subcaption}

\usepackage{amsmath}
\newcommand\numberthis{\addtocounter{equation}{1}\tag{\theequation}}

\usepackage{url}
\usepackage{hyperref}

\usepackage{multirow}

\hypersetup{colorlinks=true,breaklinks,linkcolor=blue,urlcolor=blue,citecolor=green}

\begin{document}
\title{Theory of $L$-edge spectroscopy of strongly correlated systems}

\date{\today}
 
\author{Johann L{\"u}der}
\affiliation{Department of Physics and Astronomy, Uppsala University, Box-516,Uppsala  SE-751 20 Sweden}
\affiliation{Department of Mechanical Engineering, National University of Singapore, 21 Lower Kent Ridge Rd, Singapore 119077, Singapore}

\author{Johan Sch\"ott \footnote{johan.schott@physics.uu.se}}
\email[]{johan.schott@physics.uu.se}
\affiliation{Department of Physics and Astronomy, Uppsala University, Box-516,Uppsala  SE-751 20 Sweden}

\author {Barbara Brena}
\affiliation{Department of Physics and Astronomy, Uppsala University, Box-516,Uppsala  SE-751 20 Sweden}

\author {Maurits W. Haverkort}
\affiliation{Institute for Theoretical Physics, Heidelberg University, Philosophenweg 16, D-69120 Heidelberg, Germany}

\author{Patrik Thunstr\"om}
\affiliation{Department of Physics and Astronomy, Uppsala University, Box-516,Uppsala  SE-751 20 Sweden}

\author{Olle Eriksson}
\affiliation{Department of Physics and Astronomy, Uppsala University, Box-516,Uppsala  SE-751 20 Sweden}
\affiliation{School of Science and Technology, \"Orebro University, SE-701 82 \"Orebro, Sweden}

\author {Biplab Sanyal}
\affiliation{Department of Physics and Astronomy, Uppsala University, Box-516,Uppsala  SE-751 20 Sweden}

\author{Igor Di Marco}
\affiliation{Department of Physics and Astronomy, Uppsala University, Box-516,Uppsala  SE-751 20 Sweden}

\author {Yaroslav O. Kvashnin}
\affiliation{Department of Physics and Astronomy, Uppsala University, Box-516,Uppsala  SE-751 20 Sweden}

\begin{abstract}
X-ray absorption spectroscopy measured at the $L$-edge of transition metals (TMs) is a powerful element-selective tool providing direct information about the correlation effects in the $3d$ states.
The theoretical modeling of the $2p\rightarrow3d$ excitation processes remains to be challenging for contemporary \textit{ab initio} electronic structure techniques, due to strong core-hole and multiplet effects influencing the spectra.
In this work we present a realization of the method combining the density-functional theory with multiplet ligand field theory, proposed in Haverkort \textit{et al.} [\href{https://link.aps.org/doi/10.1103/PhysRevB.85.165113}{Phys. Rev. B \textbf{85}, 165113 (2012)}].
In this approach a single-impurity Anderson model (SIAM) is constructed, with almost all parameters obtained from first principles, and then solved to obtain the spectra. 
In our implementation we adopt the language of the dynamical mean-field theory and utilize the local density of states and the hybridization function, projected onto TM $3d$ states, in order to construct the SIAM.
The developed computational scheme is applied to calculate the $L$-edge spectra for several TM monoxides.
A very good agreement between the theory and experiment is found for all studied systems. 
The effect of core-hole relaxation, hybridization discretization, possible extensions of the method as well as its limitations are discussed.
\end{abstract}


\maketitle

\section{Introduction}
The interaction of x rays with matter gives insights into structural, chemical and electronic properties of materials.
For instance, spectroscopy is used nowadays to find materials for catalysis purposes\cite{catalysts, insitu-catalysts}, hydrogen storage\cite{hydrogen,C4CP02918F}, batteries\cite{batteries} and many more.
Development of the synchrotron radiation facilities allowed to make spectroscopic analyses with an unprecedented speed and resolution.
The recently achieved increase in the experimental resolution, e.g. with the use of crystal analyzers\cite{Glatzel200565}, allowed to resolve more detailed features in the absorption spectra~\cite{PhysRevLett.111.253002}.
This advancement of experimental tools constantly challenges the existing theoretical methods aimed at calculating X-ray spectroscopy, and forces the researchers to refine the approximations they use. 

The $L$-edge x-ray absorption spectroscopy (XAS) of the transition metals (TMs) is particularly interesting and informative.
In this type of experiments one primarily probes the possible transitions from the core $2p$ to unoccupied $3d$ levels.
A variety of exotic phenomena such as charge, spin and orbital orderings\cite{PhysRevLett.80.1932}, Mott transitions\cite{PhysRevLett.23.1384,RevModPhys.70.1039}, Kondo resonances, multiferroicity\cite{PhysRevB.71.224425} and superconductivity\cite{doi:10.1021/ja800073m} originate from the correlation effects in the $3d$ states.
Thus, the $3d$ states play a decisive role in defining many important properties of the TM compounds and the $L$-edge XAS directly provides their fingerprints.
Another serious advantage of $L$-edge XAS is the applicability of the so-called sum-rules~\cite{PhysRevLett.68.1943,PhysRevLett.70.694}.
This makes it possible to extract the element-resolved spin and orbital moments in ferromagnetic materials. 

As to the theoretical modeling of XAS, initially there were two main groups of methods: the ones based on density functional theory (DFT)~\cite{RevModPhys.71.1253} and those utilizing atomic multiplet theory\cite{cowan-multiplets}. 
For a review of both classes of methods, see Refs.~\onlinecite{deGroot-book,yaresko-book}.
The methods based on DFT are in principle parameter-free and provide a very detailed description of the chemical structure. 
On the other hand, they drastically simplify the electron-electron correlations and therefore work best for itinerant electron systems.
The atomic multiplet theory is completely the opposite, as it is intended to describe exactly all many-body interactions within an isolated ion.
The theory best describes very localised electronic states, like $4f$ orbitals of rare-earth elements.
All effects like crystal field splitting and hybridisation have to be taken into account by introducing \textit{ad hoc} parameters, which resulted in the development of the multiplet ligand field theory (MLFT)\cite{MLFT}.
The main drawback of the latter is an ambiguity in the choice of the parameters, which can drastically affect the final results (see, e.g., Ref.~\cite{1742-6596-190-1-012143}).

This work concerns the study of TM monoxides, since they exhibit a correlation-driven insulating state and have been extensively studied both theoretically and experimentally~\cite{RevModPhys.70.1039,doi:10.1080/00018737700101443,PhysRevLett.55.418,MnO-TMO_magnetic_moment_collapse,PhysRevB.74.195114,PhysRevLett.109.186401}.
In order to be able to calculate the $L$-edges of TMs in these systems, a method which takes into account both multiplet and band-structure effects is required.
There are three main reasons for that.
First of all, the crystal field strength is estimated to be of the order of few eVs\cite{PhysRevB.40.5715} and the metal-oxygen hybridization is appreciable, giving rise, for example, to the super-exchange interactions, that lead to a magnetic order at low temperature~\cite{PhysRevB.80.014408}.
Second, even the description of the ground state of TM oxides (TMO) using conventional first-principles DFT-based methods remains a problem, which manifests itself in the wrongly predicted metallic character.
The complexity of TMOs lays in the fact that the $3d$ orbitals are neither completely localised (like the $4f$ states in rare-earth metals), nor itinerant (like $sp$-states of Al).
In order to capture both atomistic and band characters, several methods accounting for strong on-site correlation effects have been proposed, such as DFT+$U$\cite{PhysRevB.44.943} and DFT plus dynamical mean field theory (DFT+DMFT)\cite{RevModPhys.78.865}.
Since DFT+DMFT takes into account the multiplet effects, it was shown to provide a good description of valence band spectra of TMOs~\cite{PhysRevB.74.195114,PhysRevLett.109.186401}.
The final reason why correlations are important in the XAS process is that the $L$-edge excitation involves the presence of a $2p$ core-hole in the final state.
This creates an additional attractive potential for the valence electrons, which tends to further localise the valence states.
In addition, the created core-hole has a certain symmetry, which applies additional restrictions on the allowed transitions, giving rise to very distinct multiplet features in the XAS\cite{DEGROOT1993111}.

Several attempts to include all above-mentioned effects within a single computational framework for calculating $L$-edges of TM's have been suggested.
The state-of-the-art methods include time-dependent DFT~\cite{PhysRevLett.80.4586,PhysRevB.67.115120}, multiple scattering\cite{Nesvizhskii:nl4402}, DFT+DMFT with the final-state approximation~\cite{PhysRevB.84.115102}, configuration interaction\cite{ikeno-ci} and Bethe-Salpeter equation-based~\cite{PhysRevB.82.205104,PhysRevB.83.115106} methods.
In spite of theoretical complexities, that often require heavy computational efforts, most of the methods do not deliver a sufficiently good description of XAS of TMOs.
An efficient computational scheme combining DFT and MLFT has been proposed a few years ago in Ref.~\cite{PhysRevB.85.165113}.
The method is based on construction of a compact tight-binding description of the DFT band structure by using a projection onto Wannier functions.
As a second step, this information is used to parametrize the single-impurity Anderson model (SIAM)~\cite{PhysRev.124.41}, which is the core of MLFT.
The obtained Wannier orbitals are also explicitly used to calculate the onsite electron-electron Coulomb interactions. 
In Ref.~\cite{PhysRevB.85.165113} the suggested approach was applied to several TMOs and showed systematically good agreement between theory and experiment.

In this study we adopt an alternative realization of the DFT+MLFT method by using a different set of localized orbitals and concepts from DMFT.
We present the details of the implementation and apply the developed machinery on the series of TM monoxides: MnO, FeO, CoO and NiO.
The theoretical spectra are evaluated over a parameter space defined by the relevant interactions of these compounds, and a detailed comparison is made to experimental results. 

\begin{figure}
\centering
\includegraphics[width=0.97\columnwidth]{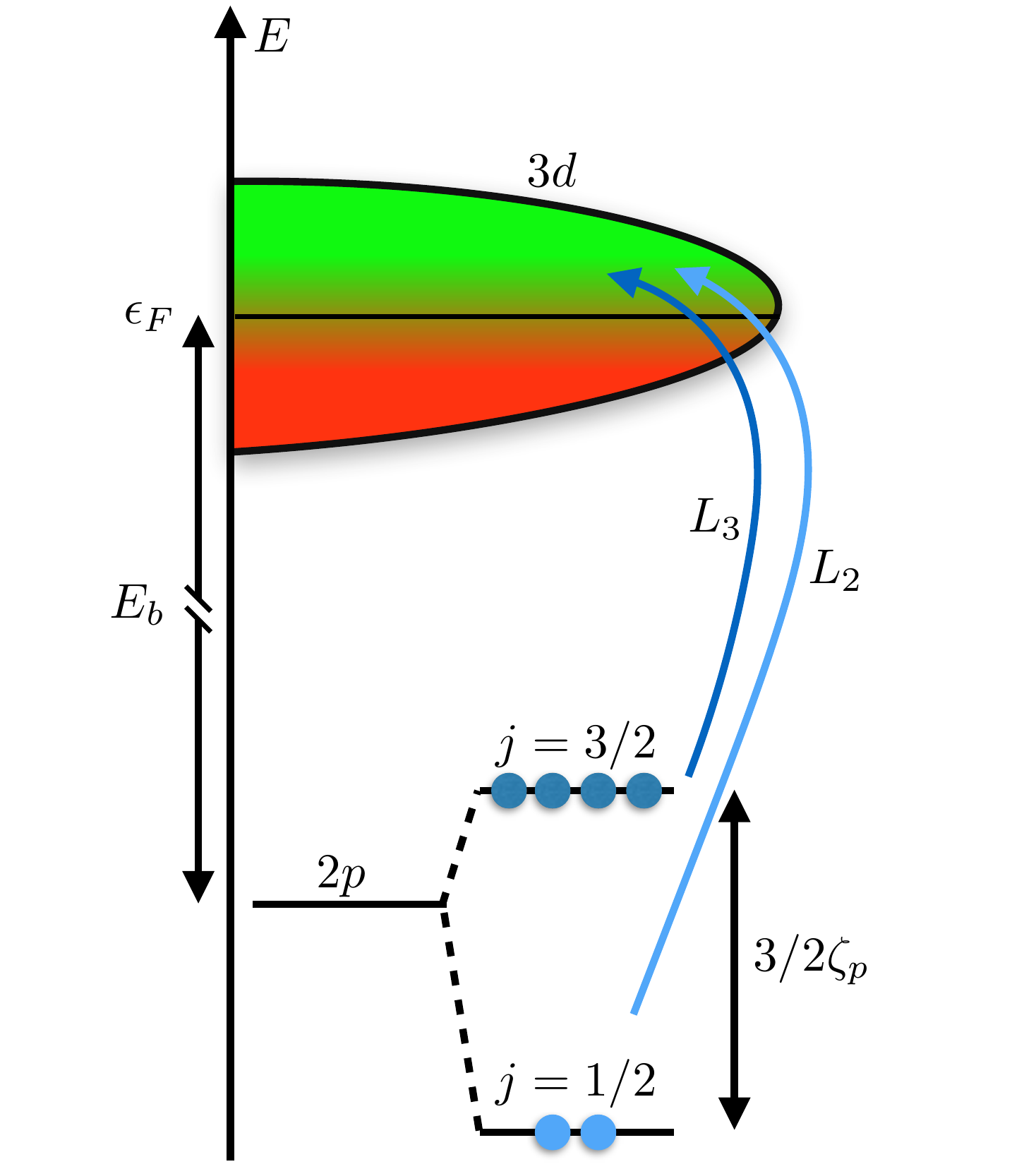}
\caption{
Schematic picture of the $L_{2,3}$-edge XA process. An electron is excited from the TM-2p core states to the TM-3d valence states. Note that the energy scale is schematic. In the presently investigated compounds the binding energy $E_b$ of the $2p$ state is of the order of 700 eV, the spin-orbit splitting $\frac{3}{2}\zeta_p$ of the $2p$ level is of the order of 10 eV and the width of the unoccupied part of the $3d$ valence band is a few eV.
}
\label{pic:XAS_schematic}
\end{figure}

\section{Theoretical aspects \label{sec:theory}}

In the following, we describe a combined DFT+MLFT approach to compute x-ray absorption (XA) spectra. 
The basic process leading to the XA $L_{2,3}$ spectra is schematically depicted in Fig.~\ref{pic:XAS_schematic}, in which an incoming photon excites an electron from the TM-$2p$ core states to the TM-$3d$ valence states. 
The aim of this section is to treat this process in a proper theoretical framework, and will result into the final expression for XA spectrum in Eq.~\eqref{eq:XAS}.  

In principle, if one had the exact density functional describing the electron-electron interaction, it would not be possible to calculate XA spectra since Kohn-Sham quasiparticles, obtained in DFT, are only meant to describe the ground state. 
Furthermore, in reality, an approximate exchange-correlation functional has to be employed.
The local density approximation (LDA) is derived for a density of a homogenous electron gas, giving rise to results, if applied to complex materials, of approximate character.
In the approach presented in this work, we instead treat the Coulomb interaction on a many-body level within MLFT, by solving the SIAM for valence $3d$ electrons.
All other quantities, such as crystal field and hybridisation are assumed to be well-described by DFT. 
The solution of the SIAM is used to obtain the excitation spectrum.

Our theory is a generalization of the DFT+MLFT presented in Ref.~\cite{PhysRevB.85.165113} by using the language of DMFT.
This allows us to go beyond the cluster model, where the TM $3d$ orbitals hybridize only with nearest-neighbor ligands~\cite{PhysRevB.96.045111}.
The formulated approach is rather general and does not depend on the particular implementations of its steps. 
Nevertheless, the calculated XA spectra will depend on the choice of local orbitals used in the SIAM.
This issue is however present for all first-principle methods based on the construction of a set of localized orbitals, e.g. DFT+U, DFT+DMFT and many other, since there is no unique definition for those in a solid~\cite{RevModPhys.78.865}.

\subsection{One-particle Hamiltonian and the hybridization function}
In this section, we explain how the single-particle (i.e. non-interacting) Hamiltonian of the impurity and bath orbitals are constructed from DFT results.

The one-particle Green's function of the lattice encodes the ligand-field contribution and the crystal-field splitting of the TM $3d$ orbitals, 
and is defined as the resolvent of the lattice-momentum dependent Hamiltonian $\hat{h}_k^\text{DFT}$:
\begin{eqnarray}
\hat g_{k,0}(\omega)  = ((\omega + \mu)\hat{1} - \hat{h}_k^\text{DFT})^{-1},
\end{eqnarray}
where $\mu$ is the chemical potential and $\hat{1}$ is the identity operator.
The local Green's function of the impurity orbitals situated at site $R$ is constructed by projecting $g_{k,0}(\omega)$ on the set of selected impurity orbitals:
\begin{eqnarray}
\hat G_{R,0}(\omega) = \sum_k \hat P_{R,k} \hat g_{k,0}(\omega) \hat P_{R,k},
\label{eq:Glocal}
\end{eqnarray}
where $\hat P_{R,k}$ is the projection operator from lattice momentum $k$ to orthonormal orbitals on the impurity site $R$. 
More details about the choice of projection will be given in the next section.
The hybridization function 
\begin{eqnarray}
\hat \Delta_{R}(\omega) = (\omega +\mu)\hat{1} -\hat G_{R,0}(\omega)^{-1} - \hat H_R
\label{eq:hyb_operator}
\end{eqnarray}
gives information at which energies and how strong the impurity interacts with its surrounding.
In Eq.~\eqref{eq:hyb_operator} the local Hamiltonian $\hat{H}_R$ is calculated from $\hat{h}_k^\text{DFT}$ by using the same projection as in Eq.~\eqref{eq:Glocal}.
Formally one can write the operator in Eq.~\eqref{eq:hyb_operator} in matrix form
\begin{eqnarray}
\Delta_{R,dd}(\omega) = \sum_b \frac{|V_{bd}|^2}{\omega - \epsilon_b},
\label{eq:hyb_sum}
\end{eqnarray}
where $V_{bd}$ are the impurity-bath hopping parameters and $\epsilon_b$ are the positions of the bath orbitals.
In a generic system the sum over bath states in Eq.~\eqref{eq:hyb_sum} is infinite. However, for localized impurity states, the imaginary part of the hybridization function often consists of several distinct peaks. In this case one can approximate the sum in Eq.~\eqref{eq:hyb_sum} by including only a finite (and usually small) number of bath states. This approximation is routinely done in the exact diagonalization (ED) solver in DMFT~\cite{PhysRevLett.72.1545}. This name stems from the fact that the finite-size SIAM can be exactly solved by direct diagonalization of the full many-body Hamiltonian, or at least its relevant sectors.    
In the presented technique, the projection is performed only on the TM $3d$ orbitals, which are relatively well-localised in TMOs.
The influence from more delocalized O-$2p$ states to the XAS process is incorporated in the hybridization function for the TM $3d$ orbitals.

\subsection{Multiplet theory}
The single-particle Hamiltonian, obtained from DFT, can be combined with Coulomb interaction terms and the resulting many-body Hamiltonian (including the core $2p$-states) corresponds to a SIAM of the form:
\begin{align*}
\hat H &= \sum_{i j}   \epsilon_{d_{i,j}} \hat{d}^{\dagger}_{i} \hat{d}_{j} + \sum_{i}  \left (   \epsilon_{b_i} \hat{b}_i^{\dagger} \hat{b}_i + \sum_{j} V_{i,j} (\hat{d}_{j}^\dagger \hat{b}_i  + \text{h.c.} )    \right )     \\ 
&  + \zeta_{d} \sum_{i j} \bra{d_i} \vec{\hat{l}} \cdot \vec{\hat{s}}\ket{d_j}\hat{d}^{\dagger}_i \hat{d}_j   + \sum_{i j k l}  {U}^{dd}_{ijkl} \hat d^{\dagger}_i\hat d^{\dagger}_j\hat d_l\hat d_k \\
& + \sum_i \epsilon_p \hat{p}_i^\dagger \hat{p}_i + \zeta_p \sum_{i j} \bra{p_i} \vec{\hat{l}} \cdot \vec{\hat{s}}\ket{p_j}\hat{p}^{\dagger}_i \hat{p}_j   \\
& + \sum_{i j k l}  {U}^{pd}_{ijkl} \hat d^{\dagger}_i\hat p^{\dagger}_j\hat p_l\hat d_k,
\numberthis \label{eq-siam}
\end{align*}
where the annihilation operators $\hat{d}_i$, $\hat{b}_i$ and $\hat{p}_i$ respectively remove an electron from a $3d$, a bath and a $2p$-core spin-orbital state.
The super-indices $i, j, k, l$ run over for all spin-orbitals within the $3d$-shell, the bath or the $2p$-core. 
The non-relativistic single-particle energies are $\epsilon_{d_{i,j}}$, $\epsilon_{b_i}$ and $\epsilon_p$.
The spin-orbit coupling of the $3d$ ($2p$) states is described in Eq.~\eqref{eq-siam} by the coupling-constant $\zeta_d$ ($\zeta_p$), the angular momentum operator $\vec {\hat{l}}$ and the spin operator $\vec {\hat{s}}$. The strong spin-orbit coupling for the $2p$-states results in the splitting of the $L_3$ and $L_2$-edges in the XA spectrum.
The on-site Coulomb repulsion between the $3d$-electrons is described by the $U^{dd}$ tensor, which can be expressed via Slater-Condon integrals $F_{dd}^0$, $F_{dd}^2$ and $F_{dd}^4$, and 
the Coulomb interaction between the $2p$-core hole and $3d$ electrons is described by the interaction tensor $U^{pd}$, which can be expressed via Slater-Condon integrals $F_{pd}^0$, $F_{pd}^2$,  $G_{pd}^1$ and $G_{pd}^3$. 
Slater-Condon integrals are special cases of the more general equation
\begin{align*}
R^k(n_1 l_1,n_2 l_2,n_3 l_3,n_4 l_4) & = \int_0^{\infty} dr r^2 \int_0^\infty dr' r'^2  R_{n_1 l_1}(r) \\
& \times   R_{n_2 l_2}(r') \frac{r_<^k}{r_>^{k+1}} R_{n_4 l_4}(r) R_{n_3 l_3}(r'),
\numberthis
\end{align*} 
where $R_{n l}$ is the radial wave function, $n$ is the principal quantum number and $l$ the angular momentum, namely
\begin{equation}
\begin{split}
F^k(n l;n' l') &= R^k(n l, n' l',n' l',n l)  \\ 
G^k(n l;n' l') &= R^k(n l, n' l',n l,n' l').
\end{split}
\end{equation}
In a cubic harmonics basis the non-spin polarized energies $\epsilon_{d_{i,j}}$, $\epsilon_{b_i}$ are reduced to $\epsilon_{d,t}=\epsilon_d + \alpha_t 10Dq$ and $\epsilon_{b,t} = \epsilon_b + \alpha_t \delta_b$, with $t \in \{ e_g, t_{2g} \}$, $\epsilon_d$ ($\epsilon_b$) the $e_g$-$t_{2g}$ averaged $3d$ (bath $b$) energy, $10Dq$ the crystal-field splitting between the $e_g$ and $t_{2g}$, $\delta_b$ the $e_g$-$t_{2g}$ splitting of bath state $b$, $\alpha_{e_g}=\frac{3}{5}$ and $\alpha_{t_{2g}}=-\frac{2}{5}$.
In this basis also the hybridization parameter $V_{i,j}$ simplifies to $V_{b,t}$, where $V_{b,t}$ describes hopping between a $3d$ orbital and a bath $b$ orbital of character $t$. Appendix~\ref{appendix:hyb} describes how $V_{b,t}$ is obtained via a fitting to the hybridization function.

The solution of  Eq.~\eqref{eq-siam} results in the set of many-body eigenstates $\ket{i}$, each expressed through a sum of Slater determinants, and the corresponding eigenenergies $E_i$:
\begin{eqnarray}
\hat H  \ket{i} = E_i \ket{i}
\label{eq-Ei}
\end{eqnarray}
Once the $\ket{i}$ corresponding to the few lowest energies are found, all statistical properties such as occupation numbers and spin moments can be directly obtained.
The XA intensity is computed from  
\begin{align*}
I (\omega) =  \frac{1}{Z}  \sum_i &  -\Im \Bigg[{\bra{i}\hat D^\dagger \frac{1}{\omega - (\hat{H}-E_i) + \mathrm{i} \Gamma/2 } \hat D \ket{i}} \Bigg] \\
& \times \exp(-\beta E_i),
\numberthis
\label{eq:XAS}
\end{align*}
where the dipole operator $\hat{D}= \epsilon \cdot \hat{r}$ describes the excitation of a $2p$-core electron to the 3d-shell, with $\epsilon$ being the light polarization, $\hat{r}$ the position operator, $\Gamma$ the imaginary offset from the real axis which gives a Lorentzian broadening of the spectra, $Z$ the partition function, and $\beta$ the inverse temperature~\cite{DEGROOT1994529,Haverkort_thesis}.

\section{Computational details \label{sec:computational_details}}

The isotropic XAS calculations were conducted for NiO, CoO, FeO and MnO. 
These TM oxides all have the same rock-salt crystal structure and experimental lattice parameters have been used~\cite{PhysRevB.80.014408}. 
Self-consistent non-spin polarized DFT calculations were performed with a 26$\times$26$\times$26 k-point mesh sampling the Brillouin zone. 
We used a linear muffin-tin orbital method (LMTO) with a full-potential as well as a warped LDA potential (see below) as implemented in the "RSPt" code\cite{rspt-book,rspt-web} to solve the DFT problem.
The set of localized impurity orbitals is constructed by projecting the total electron density on a set of L{\"o}wdin orthogonalized LMTOs for the TM $3d$ orbitals, denoted as "ORT" in Ref.~\cite{PhysRevB.76.035107,PhysRevLett.109.186401}.
The discretization of the hybridization function is described in Appendix~\ref{appendix:hyb}.

Most of the Slater-Condon integrals are calculated using the projected $3d$ wave functions.
However, the screened values of $F_{dd}^0$ and $F_{pd}^0$ are difficult to calculate and are treated as tuneable parameters.
It is worth mentioning that methods which allow to estimate their values from complementary experimental techniques exist~\cite{PhysRevLett.76.4215}.
From the theory side, several methods for calculating the screened value of $F_{dd}^0$ from first-principles have been proposed~\cite{PhysRevB.43.7570, PhysRevB.71.035105, PhysRevB.70.195104}.
However, one should bear in mind that the screened values of both $F_{dd}^0$ and $F_{pd}^0$ depend on the choice of the projected low-energy subspace and of the correlated orbitals, and are therefore not directly transferable from one code to another~\cite{PhysRevB.86.165105}. On the other side, since the XA involves a charge-neutral excitation, the spectrum is not very sensitive to $F_{dd}^0$, $F_{pd}^0$~\cite{PhysRevB.85.165113}.

Another important aspect is the double counting (DC) correction, which has to be subtracted from the DFT-derived Hamiltonian. 
This is done in order to remove the contribution of the Coulomb repulsion that is already taken into account at the DFT level. 
The DC correction is not uniquely defined and its choice is known to influence the DFT+U and DFT+DMFT results~\cite{Karolak201011}.
In this work, we apply a DC that is normally used in MLFT by considering the relative energy for different configurations. 
The charge-transfer energy is the energy difference between configurations $d^{n_d+1}\underbar{b}$ and $d^{n_d}$ and can be expressed as  
$\Delta_b^\mathrm{CT}  = \epsilon_d^{(0)} - \epsilon_b + \delta_\mathrm{CT}$~\cite{PhysRevB.87.205139}, where $\epsilon_d^{(0)}$ is the on-site $3d$ energy before the double-counting correction. In this work, the charge transfer energy correction $\delta_{\text{CT}}$ is treated as a parameter. 
For all four studied systems we use $\delta_{\text{CT}}=1.5$ eV but we remark here that the XA spectra are not very sensitive to $\delta_{\text{CT}}$. 
See Appendix~\ref{appendix:DC} for more details.

A temperature of 300 K is used in Eq.~\eqref{eq:XAS}, which is above the experimental Neel temperature for all studied systems, except for NiO~\cite{PhysRevB.80.014408}.  
The paramagnetic phase is studied by having no exchange field present in Eq.~\eqref{eq-siam}. 
The solution of the SIAM is attained using the Quanty software~\cite{quanty-web,PhysRevB.85.165113,0295-5075-108-5-57004,PhysRevB.90.085102,1742-6596-712-1-012001}. 
The basis vectors for the ground state (GS) are obtained by using a Lanczos algorithm starting with a random d$^{n_d}$ configuration, where $n_d$ is the (initial) occupation of $d$-orbitals, and generate the so-called tridiagonal Krylov basis.

\begin{table}
\caption{Summary of the charge-transfer energy correction, Slater-Condon integrals and spin-orbit coupling parameters used in the MLFT calculations. $\delta_{\text{CT}}$, $F_{pd}^0$ and $F_{dd}^0$ are treated as free parameters ($F_{dd}^0$ from Ref.~\cite{PhysRevLett.109.186401}) while the other parameters are calculated within RSPt~\cite{rspt-book,rspt-web}. Values are in eV.} 
\begin{center}
\begin{tabular}{ | c | c  c  c | c  c  c  c  c  c  c | } 
\hline
     & $\delta_{\text{CT}}$ &  F$_{pd}^0$  &   $F_{dd}^0$  & $F_{pd}^2$    & $G_{pd}^1$ & $G_{pd}^3$ & $F_{dd}^2$ & $F_{dd}^4$ &  $\zeta_{p}$  & $\zeta_{d}$  \\ \hline
MnO  &  1.5                 &   7.5        &    6.0        &  5.6          & 4.0        &   2.3      & 9.0        & 6.1        &   6.936       &      0.051      \\   
FeO  &  1.5                 &   7.5        &    6.5        &  6.0          & 4.3        &   2.4      & 9.3        & 6.2        &   8.301       &      0.064       \\
CoO  &  1.5                 &   8.0        &    7.0        &  6.4          & 4.6        &   2.6      & 9.6        & 6.4        &   9.859       &      0.079       \\              
NiO  &  1.5                 &   8.9        &    7.5        &  6.8          & 5.0        &   2.8      & 9.9        & 6.6        &   11.629      &      0.096       \\  
\hline
\end{tabular} 
\end{center}
\label{tab:ad-ve}
\end{table}

\begin{table}
\caption{Summary of the 3d occupation. The first row contains occupation used as input for the D.C. calculations presented in Sec.~\ref{sec:computational_details}, and the following rows contain occupations obtained from solving Eq.~\eqref{eq-Ei}. } 
\begin{center}
\begin{tabular}{| c | c  c  c  c |} \hline
            &  MnO     &  FeO     &   CoO    &    NiO     \\ \hline
$n_d$  &   5  & 6 & 7 & 8 \\ \hline
$n_d^\text{calc.}$, 0 bath & 5 & 6 & 7  & 8 \\ 
$n_d^\text{calc.}$, 1 bath & 5.116 & 6.179 & 7.187 & 8.194 \\
$n_d^\text{calc.}$, 2 bath & 5.118 & 6.175 & 7.176 & 8.176 \\
$n_d^\text{calc.}$, 3 bath & 5.144 & 6.209 & 7.197 & 8.177 \\ \hline
\end{tabular} 
\end{center}
\label{tab:occupation}
\end{table}

\section{Results and Discussion}
Slater-Condon integrals, charge-transfer energy correction and spin-orbit couplings are summarized in Table~\ref{tab:ad-ve}, impurity occupation (both initial guess and calculated value) are summarized in Table~\ref{tab:occupation}, hybridization parameters are summarized in Table~\ref{tab:bath_LDA} and on-site energies in Table~\ref{tab:on-site-energy}.  
The calculated Slater-Condon values of $F_{dd}^2$ and $F_{dd}^4$ have been multiplied with screening factors 0.82 and 0.88, respectively~\cite{PhysRevB.79.165104}, while the calculated unscreened values of $F_{pd}^2$, $G_{pd}^1$, $G_{pd}^3$ are used.
In this work, the only parameters relevant for the XAS process that are not calculated from DFT are $F_{pd}^0$, $F_{dd}^0$ and $\delta_{\text{CT}}$.   
The eigenstates in Eq.~\eqref{eq-Ei} are superpositions of different configurations, e.g. d$^{n_d}$, d$^{n_d+1}$\underline{b} and d$^{n_d+2}$\underline{b$^2$}, where $\underline{b}$ represents a ligand hole. On top of this, also the temperature average will mix occupation numbers.
This results in an effective occupation number $n_d^\text{calc.}$, which is slightly higher than $n_d$, see Table~\ref{tab:occupation}. 

The XA spectra of the TMOs are computed according to Eq.~\ref{eq:XAS}. The effective broadening parameter $\Gamma$, which gives a uniform Lorentzian broadening of the spectra, was set to $0.4$ eV. 
This broadening is a simplification of the more complex mechanisms leading to the broadening observed in experiments, e.g. $L_2$ is broader than $L_3$ due to its shorter core-hole lifetime, primarily due to Coster-Kronig decay~\cite{0305-4608-11-8-025}. 
However, we use a constant broadening of the theoretical spectra. 
To facilitate comparison with the experimental spectra, all theoretical spectra are shifted in energy.

\begin{figure*}
\centering
\includegraphics[]{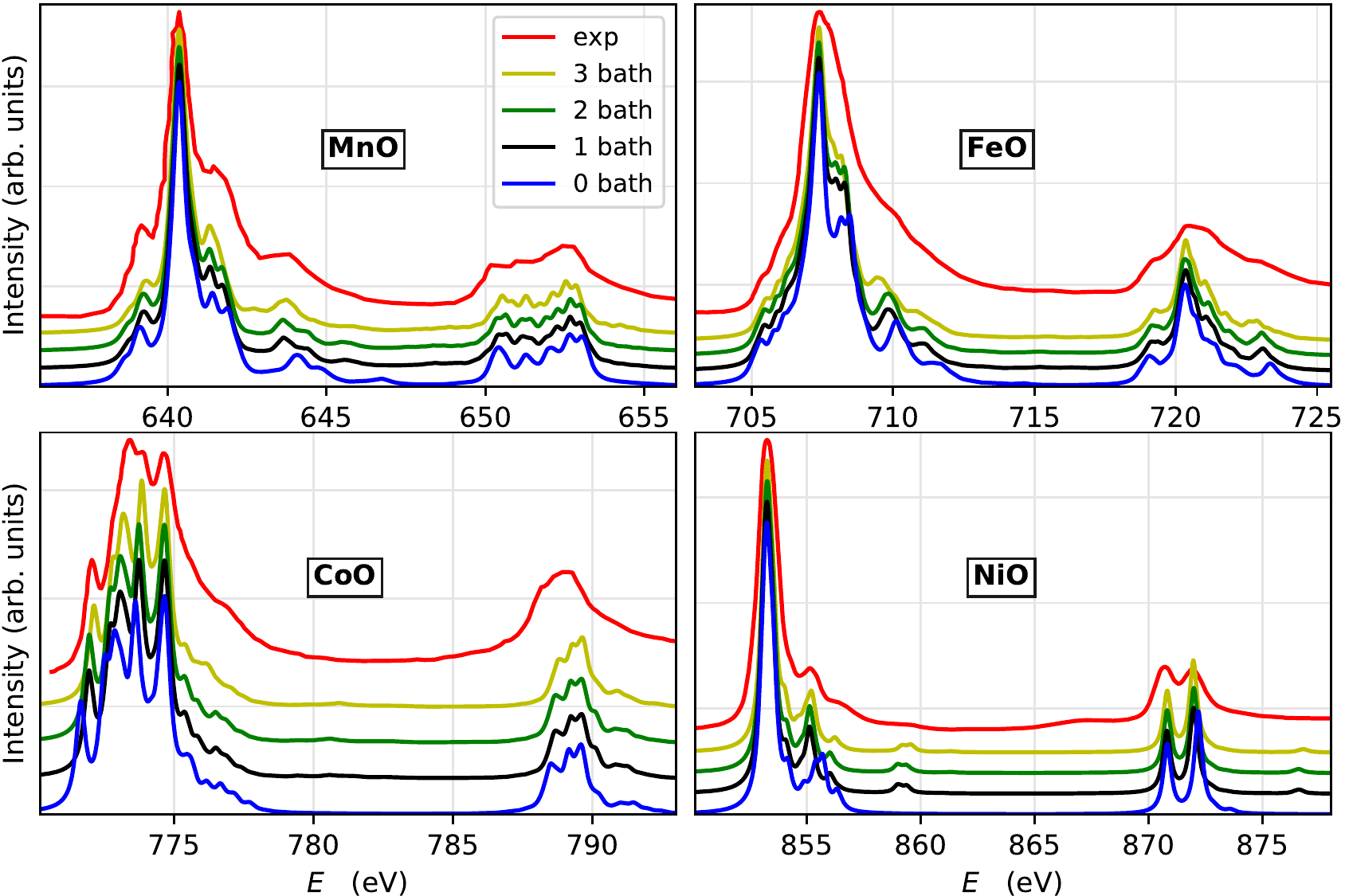}
\caption{(Color online) Comparison between the computed and the experimental XA spectra (red curves) of MnO~\cite{doi:10.1021/jp021493s}, FeO~\cite{PhysRevB.64.214422}, CoO~\cite{0953-8984-5-14-023} and NiO~\cite{PhysRevB.57.11623}. A different number of bath states are considered in the theoretical spectra.}
\label{fig:compare_with_exp}
\end{figure*}

In Fig.~\ref{fig:compare_with_exp}, the theoretical spectra are compared to XAS measurements~\cite{doi:10.1021/jp021493s,PhysRevB.64.214422,0953-8984-5-14-023,PhysRevB.57.11623}.
Note that Fig.~\ref{fig:compare_with_exp} contains theoretical information of various degrees of accuracy, as shown by the curves with different number of bath orbitals, obtained from the fitting of the hybridization function $\Delta(\omega)$ (see Appendix~\ref{appendix:hyb} for more details). 
The results in Fig.~\ref{fig:compare_with_exp} clearly show that all the experimental features are basically reproduced if one bath orbital per impurity orbital is employed. 
The addition of more bath orbitals does not change the overall pictures and mostly redistributes the intensities and shifts certain features in the final XAS results. 
We will discuss the convergence in detail below, but first we make a comparison between the higher level of theory (three bath states in Fig.~\ref{fig:compare_with_exp}) to experimental observations.
\newline
\textit{NiO} $-$ 
The resemblance between the XA spectrum and the theoretical data of NiO is very good. 
The branching ratio and the peak positions are reproduced with high accuracy by the calculation. 
The experimental intensity around 867 eV can be ascribed to excitations of the $2p_{3/2}$ electrons to free-electron-like conduction states~\cite{PhysRevB.85.165113}, which were ignored in the calculated curve.
The two $L_2$ peaks have similar intensities both in experiments and in the theory and it is known including an exchange field in the theory will further improve the relative intensity~\cite{PhysRevB.57.11623}. 
\newline
\textit{FeO} $-$ 
A very good agreement is obtained for FeO. 
The main $L_3$ peak with its spread out shoulder at energies higher than 710 eV is well reproduced and the three-peak structure in the L$_2$-edge shows good similarities to the measurement. 
Regardless of the background contributions to the measured spectrum, the computed branching ratio matches well to the experimental one.
FeO is prone to be off-stoichiometric~\cite{TANNHAUSER196225,doi:10.1063/1.325770}, which leads to point and cluster defects. 
Possible off-stoichiometry present in the actual samples may further contribute to the remaining differences between theory and experiment.
\newline
\textit{MnO} $-$ 
The $L_{2,3}$ XAS line profile for MnO has rich features. 
The $L_3$ edge contains three distinct peaks and a wide high-energy shoulder region. 
The L$_2$ edge is broad with no main peak visible. 
Also for this case, the presented approach resolves most of the details of the observed spectral shape.
We can see indications of a slight underestimation of the crystal-field splitting provided by our approach.
\newline
\textit{CoO} $-$ 
A less pronounced agreement is found for the $L_3$ edge in CoO. 
The main peak structure is well reproduced and the agreement with experiment for the pre-peak position improves with increasing number of bath states used. 
The shoulder region between 775 and 779 eV is underestimated.  
This underestimation could be a consequence of charging effects in the experimental data. See spectra of thin CoO films and CoO mixed with Ag in Ref.~\cite{Haverkort_thesis}.
The position of the $L_2$-edge for CoO does not meet the quantitative similarity of the other TMOs' $L_2$ sub-edge. This could be related to the interplay of spin-orbit coupling and local non-cubic distortions not included in the current calculations. These effects are known to be more important for CoO than for the other studied TMOs~\cite{PhysRevLett.95.187205,Haverkort_thesis}. It is also established that the XA spectrum for CoO is sensitive to the temperature~\cite{DEGROOT1994529,PhysRevLett.95.187205,Haverkort_thesis}.  

\begin{figure}
\centering
\includegraphics[]{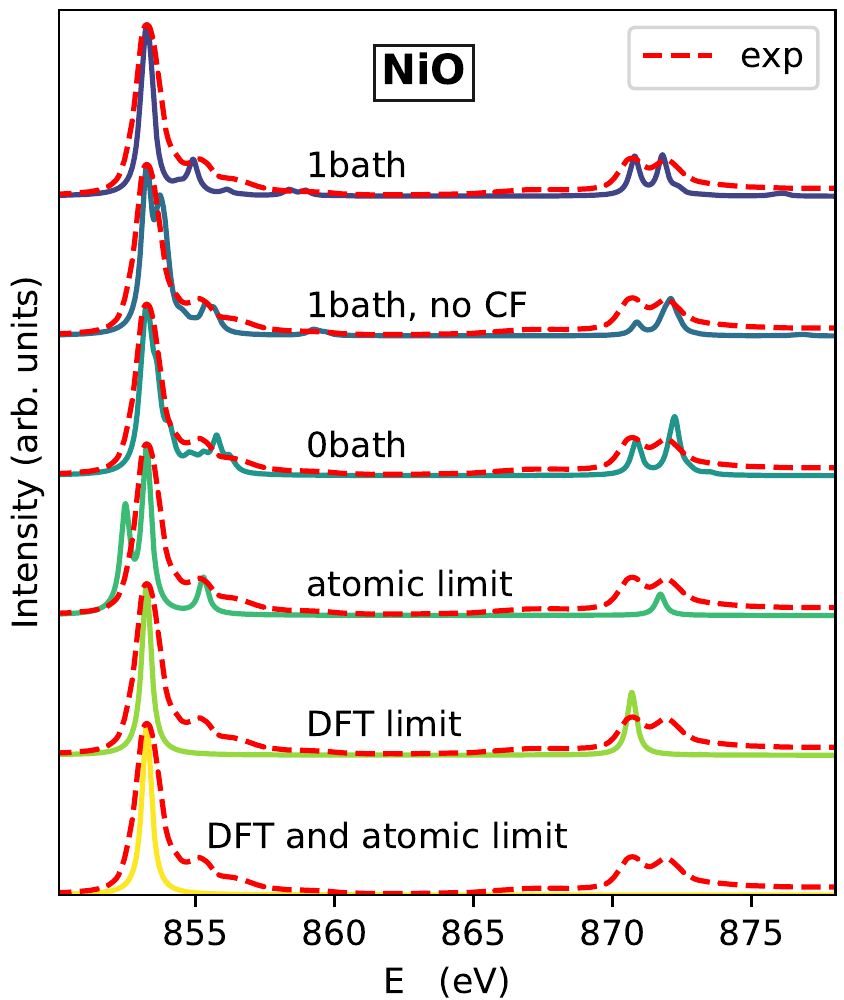}
\caption{(Color online) Calculated XA spectra of NiO. Various terms in the Hamiltonian in Eq.~\eqref{eq-siam} are removed. Description of the label names, from top to bottom: 1bath: all terms are present and one bath orbital per correlated orbital is used. 1bath, no CF: degenerate $e_g$ and $t_{2g}$  (both $10Dq$ and $\delta_b$ is zero). 0bath: no hopping to the environment. atomic limit: no hopping to the environment and degenerate $e_g$ and $t_{2g}$. DFT limit: All Slater-Condon integrals are set to zero.}
\label{fig:remove_terms_in_H}
\end{figure}

An advantage with the presented approach is the possibility to investigate the impact that separate terms in the Hamiltonian [Eq.~\eqref{eq-siam}] have on the XA spectrum. 
Figure~\ref{fig:remove_terms_in_H} illustrates as an example the NiO $L_{2,3}$ spectra obtained by neglecting selected Hamiltonian terms, e.g. crystal-field splitting, hybridization and Slater-Condon integrals.
The top panel is the spectrum obtained by including all terms in the Hamiltonian and using one bath orbital per correlated orbital. 
By removing both the energy splitting of the $e_g$ and $t_{2g}$ orbitals ($10Dq=0$) and the bath splitting ($\delta_b$=0), several changes occur. 
The $L_3$ edge gets a new peak next to the main peak,  the peak around 855 eV shifts up in energy, the experimental shoulder around 856 eV is absent, and the relative intensity ratio between the two $L_2$ peaks is in worse agreement with the experimental data.
The importance of the d-electrons to dynamically interact with its environment is shown in the third panel from the top by setting the hopping strengths ($V$) to zero.     
In the $L_3$ edge a new shoulder around 854 eV arises and the peak at 855 eV is shifted up too far in energy. 
Also here, without hopping, the correct relative intensity of the two peaks in the $L_2$ edge is not captured.
A more severe approximation in shown in the fourth panel, namely, the atomic limit. 
Here, all effects from the environment are removed, thus no hopping and no crystal-field splitting are considered. 
This corresponds to combining the two approximation steps above. 
Within this approximation, only one peak exists in the $L_2$ edge. 
The small peak at 856 eV is absent and a new peak around 852 eV arises. 
In the fifth panel (DFT limit), all Slater-Condon integrals are zero (no many-body physics) and as expected, only one peak per edge is obtained. 
In the last panel the approximations in the DFT and atomic limit are combined. 
The spin-orbit coupling generates states with $j=5/2$ and $j=3/2$ for the 3d orbitals and $j=3/2$ and $j=1/2$ for the 2p orbitals. 
No core excitations to the 3d $j=3/2$ state is possible since it is already fully occupied with four electrons. 
This fact combined with the dipole selection rule $\Delta j = 0,\pm 1$ hinders core excitations from the 2p $j=1/2$ state. 
Thus only the $L_3$ edge is expected within this approximation.    
A final comment in this section is that the strength of the terms of the Hamiltonian entering Eq.~\eqref{eq:XAS} depends to some extent on the presence or not of a core hole in the 2p shell. 
We analyze the effect of a core hole in Appendix~\ref{appendix:core_hole}, and find that the influence is only marginal when it comes to the spectral properties. 

\section{Conclusions}
In summary, we present an approach to compute the $L_{2,3}$-edge XA spectra and we have applied it to $3d$ TMOs. 
The calculations rely on DFT ground state calculations and a projection to localized orbitals to obtain the projected density of states, the hybridization function, Slater-Condon integral values and spin-orbit coupling parameters, using the RSPt software~\cite{rspt-book,rspt-web,PhysRevB.76.035107}. These data are used as input parameters by the Quanty software~\cite{1742-6596-712-1-012001,quanty-web}, which is used to diagonalize the Hamiltonian and include core-hole interaction with the valence band states. The DC is formulated using the concept of charge-transfer energy. 

With this approach, the computed XA spectra for MnO, FeO, CoO and NiO agree very well with experimental data, with respect to branching ratios and line shape profiles. 
We have also investigated the sensitivity of the calculated spectra with respect to the number of bath states, and find that this sensitivity is small, but rather non-linear. 

We also address peculiarities related with the non-sphericity of the potential, which appears in full-potential realizations of the DFT~\cite{PhysRevB.85.165113}.
We have used a warped potential for the present calculations, which seems to improve the estimates of the crystal-field splitting, compared to using a full-potential. 
These results are presented in Appendix~\ref{appendix:DC}.
Furthermore, we discuss how the presence of the core hole on the DFT level influences the valence band parameters. 
Our results clearly show that the core-hole induced changes in all calculated parameters are relatively small and do not dramatically influence the simulated XA spectra. 

The approach adopted here relies on the information extracted from the first principles electronic structure, which can be obtained by different means. 
Within DFT, the choice of the functional is important. In this work we employed warped-potential LDA, assuming that it gives an adequate description of the crystal-field splitting and TM-O hybridization. To get a more accurate description of the hybridization one could also use DFT combined with DMFT including more bath states, which recently was used to calculate core-level x-ray photoemission spectra~\cite{PhysRevB.96.045111}. 
Extension of the present method for clusters containing non-local Coulomb interactions is another promising direction.

In order to get a complete first-principles theory of XA spectra, the current approach has to be augmented with an ability to calculate the charge-transfer energy and screened values of $F_{dd}^0$ and $F_{pd}^0$. 
The choice of charge-transfer correction $\delta_{\text{CT}}$ is related with the DC problem and also needs to be solved. 
The $F_{dd}^0$ can be calculated by means of various methods~\cite{PhysRevB.43.7570, PhysRevB.71.035105, PhysRevB.70.195104} and it would be very useful to generalize these methods for the evaluation of $F_{pd}^0$. Further possible improvement would be to predict the screening effects on the higher order Slater-Condon integrals. 
This paper provides a tool to predict measurable quantities (i.e., core-level XAS and photoemission spectroscopy ) 
that are sensitive to the double-counting scheme used as well as the screened value of $F_{dd}^0$. 
As such, this scheme can be used to test the accuracy of different implementations.

\begin{acknowledgments}
We acknowledge the Knut and Alice Wallenberg Foundation (KAW) (Projects No. 2013.0020 and No. 2012.0031) for financial support. B.B. and O.E. thank the Swedish Research Council (VR) and eSSENCE. Y.K and J.L. thank the EUSpec COST for the financial support of the research visit to Dresden. We also appreciate the Swedish National Infrastructure for Computing (SNIC) which has provided computing time on the clusters Abisko at Ume{\aa} University,  Triolith at Link{\"o}ping University, and Beskow at KTH, Stockholm.  
For financial support we also acknowledge the Carl Trygger Foundation, Sweden.
We would like to thank L. Nordstr\"om and T. Bj\"orkman for discussions of the muffin-tin potential.  

J.L and J.S. contributed equally to this work.

\end{acknowledgments}

\appendix

\section{Discretization of the hybridization function $\Delta(\omega)$ \label{appendix:hyb}}
The hybridization functions for the studied systems (MnO, FeO, CoO and NiO) are similar, due to the mean-field like treatment with DFT. 
The $e_g$ orbitals hybridize more than the $t_{2g}$ orbitals with their surrounding of O states. 
The bath energies are picked by inspecting the DFT hybridization function and the hopping parameters are obtained by considering the weight of the hybridization intensity in the vicinity of each bath energy. 
Table~\ref{tab:bath_LDA} contains the discretized hybridization function parameters.
The hybridization weight close to the Fermi level for the $t_{2g}$ orbitals is picked up by the third set of bath orbitals. 
Note, in the hybridization function of a truly insulating state, this weight is suppressed~\cite{PhysRevB.96.045111}.

\begin{figure}
\centering
\includegraphics[]{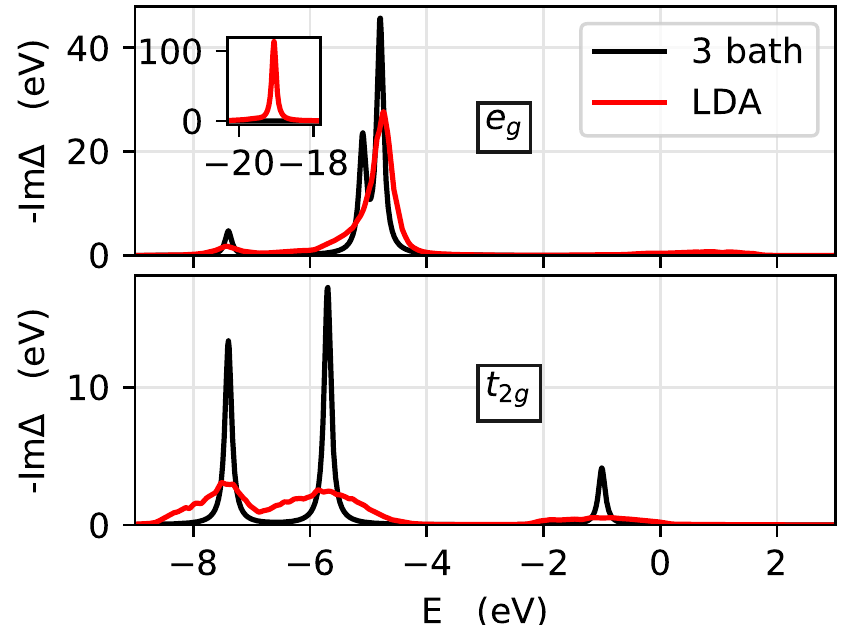}
\caption{(Color online) Comparison between the DFT hybridization function (red curves) and its fitting data using three bath states (black curves) of FeO. For sake of compactness an inset is used to show the hybridization peak corresponding to O-2s states.}
\label{fig:hyb}
\end{figure}

To compensate for the discretization approximation of the hybridization function in Eq.~\eqref{eq:hyb_sum}, we adjust the $3d$ on-site energy $\epsilon_t^{(0)}$ such that the discretized Green's function~\cite{PhysRevB.88.085112}
\begin{equation}
G_t(\omega) = (\omega-\epsilon_t^{(0)}-\sum_{b} \frac{|V_{b,t}|^2}{\omega-\epsilon_{b,t}} )^{-1},
\end{equation}
with $t \in \{ e_g, t_{2g} \}$, resemble the local DFT Green's function in Eq.~\eqref{eq:Glocal}. 
In practice, we achieved this by demanding that the imaginary part of $G_t(\omega)$ to have the same center of gravity as the imaginary part of the local DFT Green's function in a restricted energy window.
For all considered TMOs, we selected the energy window to be $[-3,3]$ eV around the Fermi level, as the electron bands with predominantly TM-$3d$ character exist in this interval.
The obtained on-site energies are presented in Table~\ref{tab:on-site-energy}. 
The DFT and the discretized hybridization function using 3 bath states per $3d$ orbital is shown in Fig.~\ref{fig:hyb} for FeO. 
For the $e_g$ orbitals, the hybridization at around -19 eV is too far away from the Fermi level to enable a charge transfer in the SIAM.
Therefore, no $e_g$ bath state is placed at around -19 eV and this hybridization is instead (implicitly) compensated by the adjustment of $\epsilon_{e_g}^{(0)}$.  

\begin{table*}
\caption{Summary of the bath parameters extracted from the hybridization functions obtained using the RSPt software~\cite{rspt-book,rspt-web}. Values are in eV. Figure~\ref{fig:hyb} shows both the DFT and the fitted hybridization function of FeO using three bath states. } 
\begin{center}
\begin{tabular}{ | c | c | c | c | c | } 
\hline
\multirow{2}{2.8em}{\#bath} &             MnO          &            FeO            &            CoO           &              NiO          \\ 
  &  $\epsilon_{b,e_g}, \epsilon_{b,t_{2g}}, V_{b,e_g}, V_{b,t_{2g}}$ & $\epsilon_{b,e_g}, \epsilon_{b,t_{2g}}, V_{b,e_g}, V_{b,t_{2g}}$  & $\epsilon_{b,e_g}, \epsilon_{b,t_{2g}}, V_{b,e_g}, V_{b,t_{2g}}$  & $\epsilon_{b,e_g}, \epsilon_{b,t_{2g}}, V_{b,e_g}, V_{b,t_{2g}}$  \\ \hline 
\multirow{1}{1em}{1}           &  -4.8, -6.7, 2.0, 1.4 & -4.8, -6.7, 2.1, 1.5 & -4.5, -6.5, 2.0, 1.4 &  -4.4, -6.5, 2.0, 1.4 \\ \hline
\multirow{2}{1em}{2}           &  -4.8, -5.7, 2.0, 1.0 & -4.8, -5.7, 2.1, 1.1 & -4.5, -5.5, 2.0, 1.0 &  -4.4, -5.5, 2.0, 1.0  \\ 
                                           &  -7.2, -7.2, 2.4, 0.9 & -7.4, -7.4, 2.5, 1.0 & -7.3, -7.3, 2.5, 0.9 & -7.3, -7.3, 2.5, 0.9  \\ \hline
\multirow{3}{1em}{3}           & -4.8, -5.7, 1.5, 1.0  & -4.8, -5.7, 1.6, 1.1 & -4.5, -5.5, 1.5, 1.0 & -4.4, -5.5, 1.6, 1.0 \\ 
                                           &  -7.2, -7.2, 2.4, 0.9 & -7.4, -7.4, 2.5, 1.0 & -7.3, -7.3, 2.5, 0.9 & -7.3, -7.3, 2.5, 0.9  \\ 
                                           &  -5.1, -1.0, 1.4, 0.6 & -5.1, -1.0, 1.3, 0.6 & -4.8, -1.2, 1.3, 0.5  & -4.7, -1.7, 1.3, 0.5 \\ 
\hline
\end{tabular}
\label{tab:bath_LDA}
\end{center}
\end{table*}

\begin{table}
\caption{On-site energies for three different bath discretizations. Values are in eV. } 
\begin{center}
\begin{tabular}{ | c | c | c | c | c |} 
\hline
\multirow{2}{2.8em}{\#bath}  &            MnO       &        FeO           & CoO                 &          NiO           \\ 
     & $\epsilon_{e_g}^{(0)}, \epsilon_{t_{2g}}^{(0)}$  & $\epsilon_{e_g}^{(0)}, \epsilon_{t_{2g}}^{(0)}$ & $\epsilon_{e_g}^{(0)}, \epsilon_{t_{2g}}^{(0)}$ & $\epsilon_{e_g}^{(0)}, \epsilon_{t_{2g}}^{(0)}$ \\ \hline
  1       &   -0.080, -0.718 & -0.232, -0.878   & -0.435, -1.046  &   -0.955, -1.560   \\ 
  2       &   -0.117, -0.734 & -0.272, -0.893   & -0.476, -1.060  &  -1.000, -1.578  \\
  3       &  -0.192, -0.742   & -0.345, -0.901   & -0.550, -1.067  &  -1.072, -1.588 \\
 \hline
\end{tabular}
\label{tab:on-site-energy}
\end{center}
\end{table}

\section{Double counting \label{appendix:DC}}
To avoid double counting the monopole part of the Coulomb interaction, we consider three configurations: $p^6 b^{10} d^{n_d}$, $p^6 b^9  d^{n_d+1}$ and $p^5 b^{10}  d^{n_d+1}$ and their corresponding energies $E_{b,0}$, $E_{b,0} + \Delta_b^\mathrm{CT}$ and $E_{b,0} + (\epsilon_d^{(0)} - \epsilon_p^{(0)})$. These energies can be expressed as~\cite{PhysRevLett.55.418} 
\begin{equation}
\label{eq:ligand_theory:core_included}
\begin{aligned}
E_{b,0} &= 6\epsilon_p+ 10 \epsilon_b +  n_d \epsilon_d \\
& + \binom{n_d}{2}  U_{dd}  + 6 n_d U_{pd},    \\
E_{b,0} + \Delta_b^\mathrm{CT}  &= 6\epsilon_p+  9 \epsilon_b + (n_d+1) \epsilon_d  \\ 
& + \binom{n_d+1}{2}  U_{dd} + 6 (n_d +1) U_{pd},  \\ 
E_{b,0} + (\epsilon_d^{(0)} - \epsilon_p^{(0)}) &= 5\epsilon_p + 10 \epsilon_b + (n_d+1) \epsilon_d  \\
& + \binom{n_d+1}{2}  U_{dd} +  5 (n_d +1) U_{pd},
\end{aligned}
\end{equation}
where $\Delta_b^{\text{CT}}$ is the charge transfer energy and $\epsilon_d^{(0)}$ ($\epsilon_p^{(0)}$) is the DFT on-site energy for the $3d$ ($2p$) states.
By solving these equations for the double-counting corrected energies $\epsilon_d$ and $\epsilon_p$, which enter in the SIAM in Eq.~\eqref{eq-siam}, and expressing the charge-transfer energy as $\Delta_b^\mathrm{CT}  = \epsilon_d^{(0)} - \epsilon_b + \delta_\mathrm{CT}$~\cite{PhysRevB.87.205139}, where $\delta_\mathrm{CT}$ is a charge-transfer correction parameter, one obtains 
\begin{equation}
\label{eq:ligand_theory:core_included:final}
\begin{aligned}
\epsilon_d   &=   \epsilon_d^{(0)} + \delta_\mathrm{CT} - n_d U_{dd} -6 U_{pd} \\       
\epsilon_p   &=   \epsilon_p^{(0)} + \delta_\mathrm{CT} - (1+n_d)U_{pd}.
\end{aligned}
\end{equation}
Note that $\epsilon_p$ only will shift the XA spectrum. 
The average Coulomb repulsion energies used in Eq.~\eqref{eq:ligand_theory:core_included} and Eq.~\eqref{eq:ligand_theory:core_included:final} are expressed in Slater-Condon integrals by
\begin{gather}
U_{dd} = F_{dd}^0 - \frac{2}{63} (F_{dd}^2+F_{dd}^4)                         \\
U_{pd}   = F_{pd}^0 - \frac{1}{15} G_{pd}^1 - \frac{3}{70} G_{pd}^3.
\end{gather}

\begin{figure}
\centering
\includegraphics[width=1.0\columnwidth]{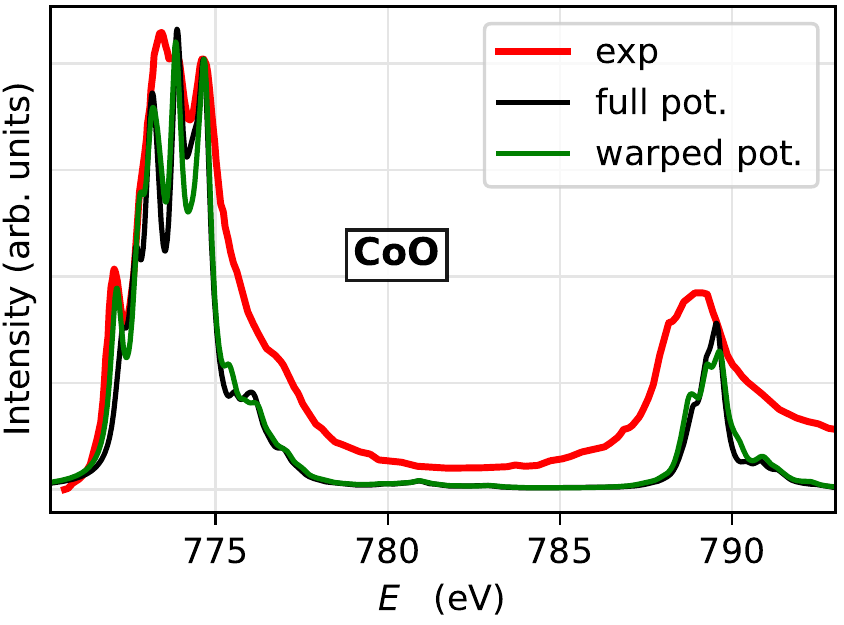}
\caption{(Color online) Comparison between the computed XA spectra using warped and full potential LDA of CoO. Experimental spectrum (red line) is shown as reference~\cite{0953-8984-5-14-023}. Three bath states are used.}
\label{fig:xas:warped_vs_full_potential}
\end{figure}

In order to avoid double counting the multipole part of the Coulomb interaction, we have used a warped LDA potential instead of a full LDA potential, as suggested in~\cite{PhysRevB.85.165113}. This means the non-spherical part of the potential inside the muffin-tin is zero. 
Removing non-spherical parts of the potential, to improve the crystal-field splitting, has been discussed in the past for $d$-electrons~\cite{PhysRevB.85.165113} in terms of double-counting and for $f$-electrons~\cite{PhysRevLett.79.2546} in terms of self-interaction. 
The main difference in the band structure between the warped and the full potential is that the $e_g$ and $t_{2g}$ bands are further apart in energy with the warped potential.
All other parameters extracted from DFT, such as
bath energies, hoppings, and Slater-Condon integrals, are barely affected by this approximation of the potential.
For comparison, the XA spectra obtained using the warped and full potential of CoO is shown in Fig.~\ref{fig:xas:warped_vs_full_potential}.
It may be seen that the differences between the two theoretical curves are small, but clearly noticeable. 
The calculation based on warped potential seems to reproduce observations better, especially the features in the low-energy region.

\section{Presence of a core-hole within DFT\label{appendix:core_hole}}
In Ref.~\cite{PhysRevB.85.165113}, the parameters of a single-particle Hamiltonian are extracted from a ground state DFT calculation with no core-hole.
The $2p$ core-hole only enters the calculation on the stage of MLFT, participating in a Coulomb interaction with the valence $3d$ states. 
Thus, the influence of a core-hole on the valence band electronic structure is not explicitly considered.
In this work we aim to quantify these so far neglected effects.

\begin{figure}
\centering
\includegraphics[width=1.0\columnwidth]{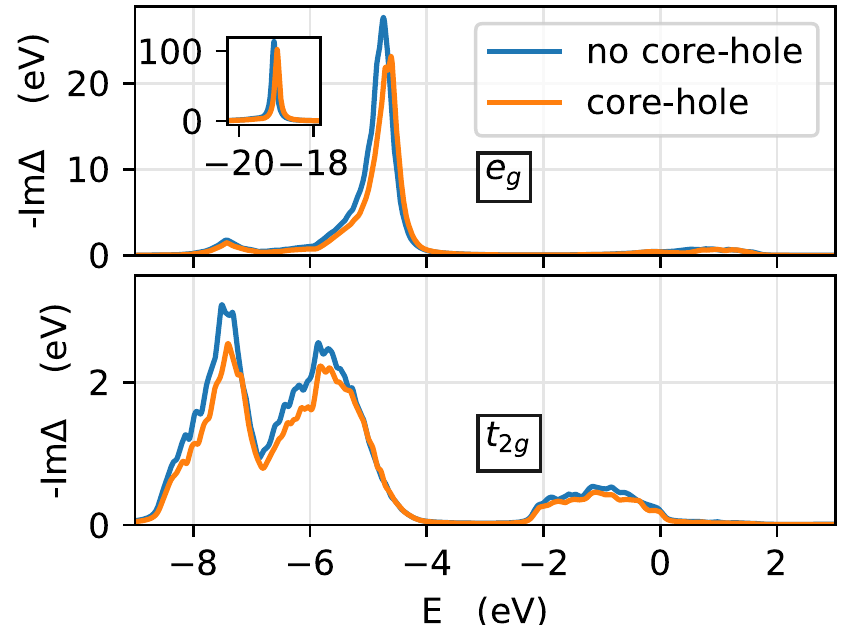}
\caption{(Color online) Hybridization function including the effect of core-hole relaxation, obtained from DFT, for FeO. }
\label{fig:hyb_core-hole}
\end{figure}

\begin{figure}
\centering
\includegraphics[width=1.0\columnwidth]{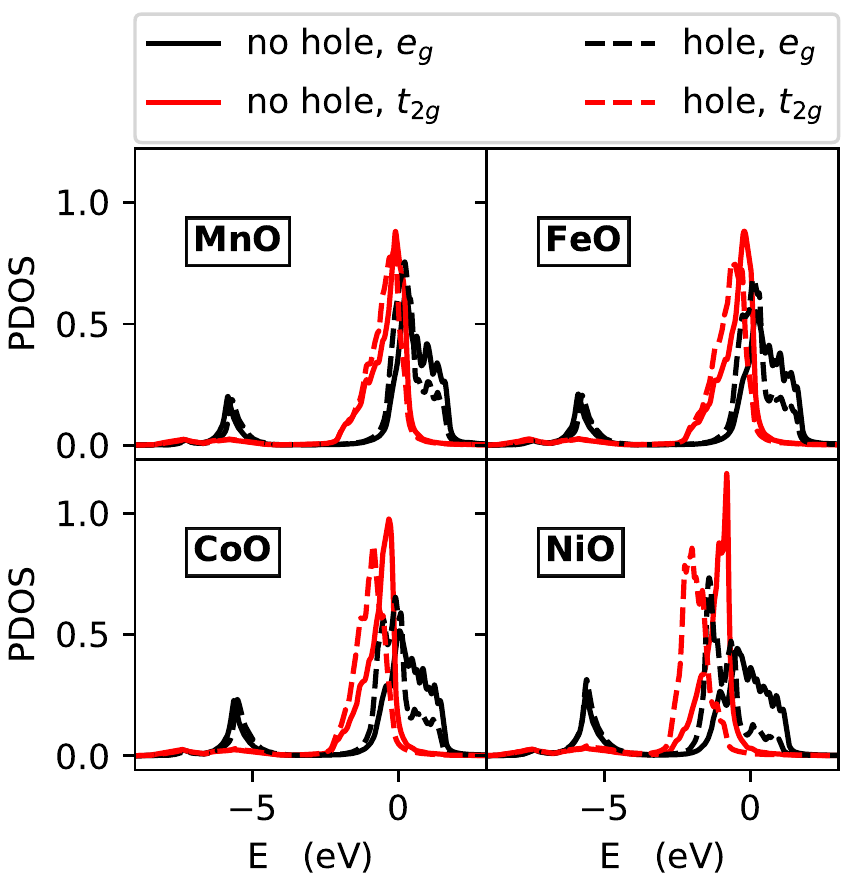}
\caption{(Color online) PDOS including the effect of core-hole relaxation, obtained from DFT.}
\label{fig:PDOS_core_hole}
\end{figure}

We have constructed a supercell and created a $2p$ core hole on one of the TM atoms. 
The created excess charge was either added as a uniform background or simply added to the valence, to maintain charge neutrality.  
Both schemes resulted in identical sets of results, which gives credence to the chosen size of the supercell.

The imaginary part of the hybridization function, with and without the core hole, is shown in Fig.~\ref{fig:hyb_core-hole} for FeO. For MnO, CoO, and NiO, the changes in the hybridization functions are similar.
The hybridization is suppressed, which roughly corresponds to a 7$\%$, 7$\%$, 8$\%$ and 9$\%$ decrease in the hopping parameters for MnO, FeO, CoO, and NiO, respectively. 
The Slater-Condon integrals $F_{pd}^2$, $G_{pd}^1$, $G_{pd}^3$, $F_{dd}^2$ and $F_{dd}^4$ are increased a few percentage points (see Table~\ref{tab:core_hole_Fdd24}).
Both findings are consistent with the idea that the core hole creates a potential, which further tends to localize the $3d$ orbitals.
The crystal-field splitting, $10Dq$, extracted from the discretized hybridization function and the PDOS in Fig.~\ref{fig:PDOS_core_hole}, changes -18$\%$, -18$\%$, -16$\%$ and -12$\%$ for MnO, FeO, CoO and NiO respectively.

\begin{figure*}
\centering
\includegraphics[width=2.0\columnwidth]{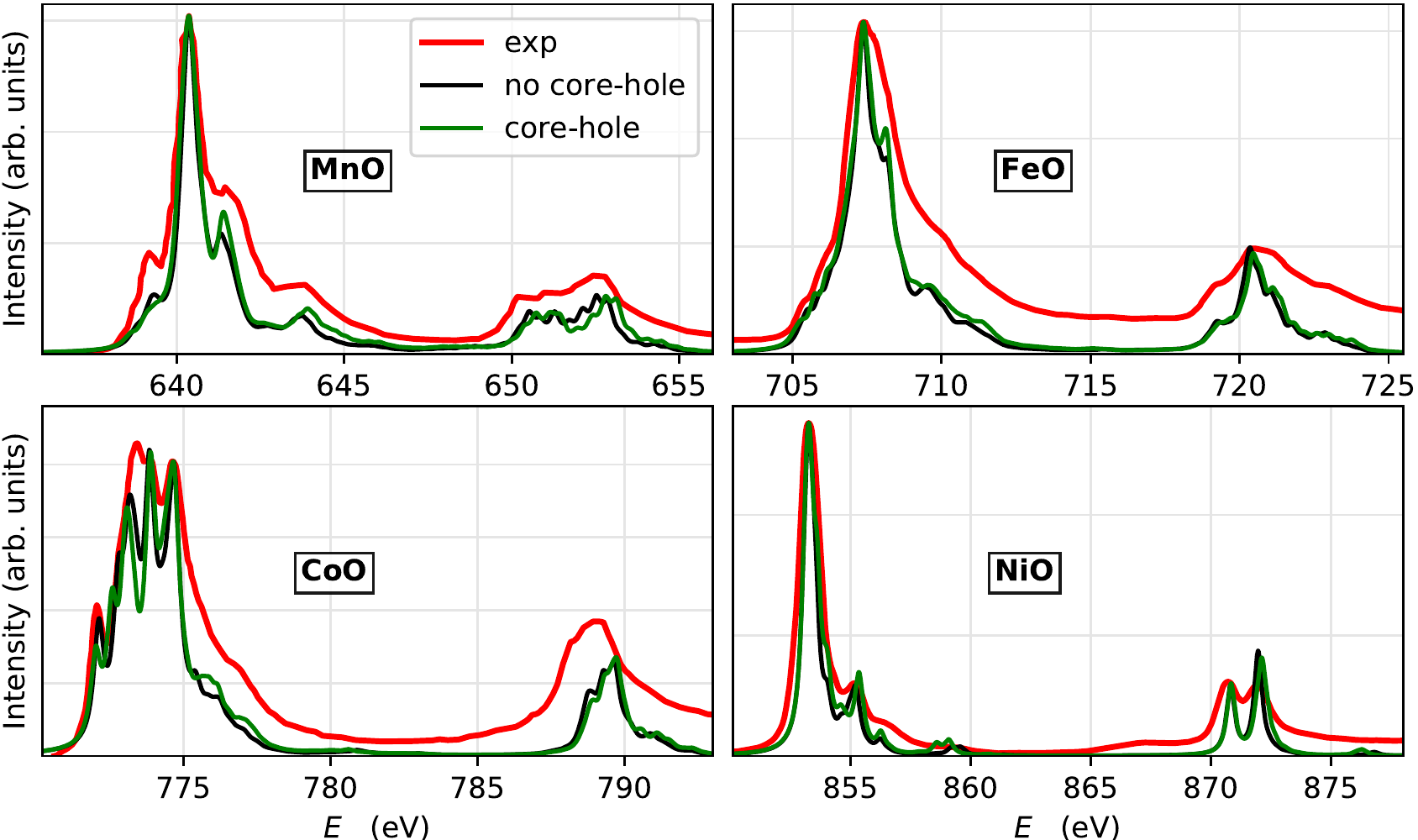}
\caption{(Color online) XAS including the effect of core-hole relaxation. One bath state is used.  }
\label{fig:XAS_core-hole}
\end{figure*}

In Fig.~\ref{fig:XAS_core-hole}, we show the simulated XA spectra taking into account the renormalization of the parameters, described above.
One can see that the resulted spectra are slightly modified with respect to the theoretical results obtained with no explicit core hole considered in DFT.
The changes in the spectra are almost entirely due to the increase of $F_{pd}^2$, $G_{pd}^1$ and $G_{pd}^3$.

\begin{table}
\caption{Relative increases of Slater-Condon integrals $F_{pd}^2$, $G_{pd}^1$, $G_{pd}^3$, $F_{dd}^2$ and $F_{dd}^4$ due to core-hole relaxation, calculated using RSPt~\cite{rspt-book,rspt-web}.} 
\begin{center}
  \begin{tabular}{  r | c c c c c }
       $\%$       & $F_{pd}^2$ & $G_{pd}^1$ & $G_{pd}^3$ & $F_{dd}^2$  & $F_{dd}^4$ \\ \hline
    MnO  &  12 & 15  & 13  & 3  & 3 \\ 
    FeO   &  10 & 14 & 17   & 3  & 3 \\ 
    CoO   &   9 &  15 &  15  & 3  & 3 \\ 
    NiO    &  10 & 12 &  14  & 4  & 3  \\
  \end{tabular}
\label{tab:core_hole_Fdd24}
\end{center}
\end{table}

\bibliography{References.bib}
\bibliographystyle{apsrev4-1}

\end{document}